\documentclass[3p]{elsarticle}
\usepackage{amsmath,amssymb,multirow,epsfig,bm,subfigure}

\newcommand{\beq}{\begin{equation}}
\newcommand{\eeq}{\end{equation}}
\newcommand{\bea}{\begin{eqnarray}}
\newcommand{\eea}{\end{eqnarray}}

\begin{document}


\title{Toward large-scale Hybrid Monte Carlo simulations of the Hubbard model on graphics processing units}

\author[OSU]{K. A. Wendt}
\ead{wendt@mps.ohio-state.edu}

\author[OSU]{J. E. Drut\corref{cor1}}
\ead{jdrut@mps.ohio-state.edu}

\author[HUT]{T. A. L\"ahde}
\ead{timo.lahde@tkk.fi}

\address[OSU]{Department of Physics, The Ohio State University, Columbus, OH 43210--1117, USA}

\address[HUT]{Helsinki Institute of Physics and Department of Applied Physics, 
Aalto University, FI-02150 Espoo, Finland}

\cortext[cor1]{Corresponding author}

\date {\today}

\begin{abstract}
The performance of the Hybrid Monte Carlo algorithm is determined by the speed of sparse matrix-vector multiplication 
within the context of preconditioned conjugate gradient iteration. We study these operations as implemented for the 
fermion matrix of the Hubbard model in $d+1$ space-time dimensions, and report a performance comparison between 
a 2.66~GHz Intel Xeon E5430 CPU and an NVIDIA Tesla C1060 GPU using double-precision arithmetic. We find speedup factors
ranging between $30-350$ for $d = 1$, and in excess of $40$ for $d = 3$. We argue that such speedups are of 
considerable impact for large-scale simulational studies of quantum many-body systems.
\end{abstract}


\maketitle


\section{Introduction}

The Hubbard model~\cite{Hubbard} and its extensions are among the most studied problems in many-body physics. 
In appropriate limits, this model is known to describe the central features of a number of systems, such as
cold atomic gases~\cite{ColdAtoms}, high-$T_c^{}$ superconductors~\cite{HighTc}, and
quantum wires and spin chains~\cite{QuantumWires}. Nevertheless, in many cases of interest the Hubbard model 
poses a non-perturbative problem, such that even qualitative studies require the use of Monte Carlo techniques. 
An example that has received significant attention in the last few years is the case of cold atoms close to a Feshbach 
resonance~\cite{UnitaryGas}, the so-called Unitary Fermi Gas~(UFG), whose universal properties are linked to multiple problems 
ranging from the crust of neutron stars~\cite{NeutronStar1,NeutronStar2} to the quark-gluon plasma created in heavy-ion collisions~\cite{QGP}.

For the above systems, the calculation of correlation functions at low-momentum or large distances is often desirable, as in the determination of 
quasiparticle spectra. This requires calculations in systems that are large enough to accommodate the long-range asymptotics of the relevant 
correlation functions. Similarly, one is often interested in the thermodynamic limit, where both the particle number $N$ and the volume $V$ are 
large. These issues are, of course, not restricted to the Hubbard model. In the case of non-perturbative quantum field theory calculations such as 
Lattice QCD~\cite{LQCD} or the low-energy effective theory of graphene~\cite{DL,Hands}, the continuum limit is attained in the vicinity of 
second-order critical points, where the relevant correlation lengths become much larger than the lattice spacing. Thus, the feasibility of such 
calculations is dependent on the availability of Monte Carlo algorithms and implementations that can handle large volumes in an efficient fashion. 
One of the most successful algorithms is referred to as Hybrid Monte Carlo (HMC). We will review the main aspects of HMC in Sec.~\ref{algorithm}. As 
we shall see, the performance of HMC is strongly dependent on the speed of matrix-vector operations and conjugate-gradient iterations, both of which 
are well suited for parallelization.

The advent of massively multithreaded computing on Graphics Processing Units~(GPUs) has generated a concerted 
effort from various scientific communities to benchmark the performance of their respective algorithms on GPU architectures.
The field of Monte Carlo 
simulations is no exception, as several groups have already set out to perform such calculations on GPU systems. Examples 
range from relatively simple test cases such as the Ising model~\cite{Demchik} to more complicated and demanding applications, 
including fixed-node Monte Carlo~\cite{Anderson} and Lattice QCD~\cite{QCDVideoGame,QCD_GPU}. In this work, we report a first study
of the feasibility of GPUs as a tool for large-scale HMC simulations of the Hubbard model, setting the stage for more complex many-body 
systems within the areas of nuclear and condensed-matter physics.

This paper is organized as follows. In Sec.~\ref{model}, we outline the model Hamiltonian and partition function. 
The CPU and GPU hardware is described in Sec.~\ref{hardware}. An overview of the Hybrid Monte Carlo~(HMC) algorithm and the implementation 
of its central components on the GPU is given in Sec.~\ref{algorithm}, followed by conclusions and a discussion of the 
observed performance in Sec.~\ref{results}.


\section{Hamiltonian and partition function \label{model}}

In this work, we shall focus on one of the simplest forms of the Hubbard model for spin-$1/2$ fermions. For a particular value of the
coupling $U$, this model represents an effective theory for fermions close to the unitary limit (such as cold atoms or dilute neutron matter). 
Nevertheless, our results should be representative of a larger class of models with similar Hamiltonians, as long as the 
dimensionality of the problem is unchanged. The Hamiltonian we shall consider is
\beq
\hat H \equiv \hat K + \hat V,
\label{HH}
\eeq
with kinetic energy
\beq
\hat K \equiv -t \sum_{\langle i,j\rangle} \sum_{s=\uparrow,\downarrow}
(\hat c^\dagger_{\sigma, i} \hat c^{}_{s,j} + \hat c^\dagger_{s,i}\hat c^{}_{s,j}),
\eeq
and potential energy
\beq
\hat V \equiv U \sum_i \hat n_{\uparrow,i}^{} \hat n_{\downarrow,i}^{}.
\eeq
Here, the indices $(i,j)$ assume values on a $d$-dimensional spatial lattice, and the notation ${\langle i,j\rangle}$ 
denotes summation over nearest-neighbor sites only. The $\hat c^\dagger_{s,i}$ and $\hat c^{}_{s,i}$ are, respectively, creation and 
annihilation operators for fermions of spin $s$ at location $i$, such that 
$\hat n_{s,i}^{} \equiv \hat c^\dagger_{s,i} \hat c^{}_{s,i}$. 

The grand canonical partition function is
\beq
\label{ZZ}
\mathcal Z \equiv \text{Tr} \, \exp(-\beta (\hat H - \mu \hat N)),
\eeq
where 
\beq
\hat N = \sum_i \sum_{s=\uparrow,\downarrow} \hat c^\dagger_{\sigma, i} \hat c^{}_{s,i}.
\eeq
Upon performing a Hubbard-Stratonovich transformation~\cite{HS,HSL} with auxiliary field $\sigma$,
Eq.~(\ref{ZZ}) can be recast in the form
\beq
\mathcal Z \, = \int \mathcal D \sigma \, (\det M[\sigma])^2 \exp(-S_g^{}[\sigma]),
\label{HSZ}
\eeq
where the fermion matrix $M$ is of the form
\beq
M \equiv 
\left( \begin{array}{ccccccc}
1 & 0 & 0 & 0 & \dots & B_{N_\tau^{}}^{}\\
-B_{1}^{} & 1 & 0 & 0 & \dots & 0 \\
0 & -B_{2}^{} & 1 & 0 &  \dots & 0 \\
\vdots & \vdots & \vdots & \vdots & \vdots & \vdots \\
0 & 0 & \dots & -B_{N_\tau^{} - 2}^{} & 1 & 0 \\
0 & 0 & \dots & 0 & -B_{N_\tau^{} - 1}^{} & 1
\end{array} \right),
\eeq
and the block matrices $B_j^{}$ are given by
\beq
\label{Bblocks}
B_j^{} \,\equiv\, \exp(-\tau K) \, \exp(-\tau V[\sigma_j^{}]),
\eeq
with the kinetic energy operator
\beq
K_{{\bf n},{\bf n}'}^{} \equiv -t \sum^{d}_{j=1} 
(\delta_{{\bf n},{\bf n}' + {\bf e}_j^{}}^{} + \delta_{{\bf n},{\bf n}' - {\bf e}_j^{}}^{}),
\eeq
and $V[\sigma_j^{}]$ the auxiliary Hubbard-Stratonovich potential, evaluated at imaginary-time slice $j$. 
The integer vectors $({\bf n}, {\bf n}')$ assume values on a $d$-dimensional spatial lattice. In $d+1$~dimensions, 
$M$ is given (up to second order in $\tau$) by
\beq
\label{dDM}
M_{{\bf n} \tau,{\bf n}'\tau'}^{} = A\,\delta_{\tau-1,\tau'}^{} \delta_{{\bf n},{\bf n}'}^{} +
\delta_{\tau,\tau'}^{} \sum_{j=1}^{d} \left( B\,\delta_{{\bf n},{\bf n}'}^{} + C\,\delta_{{\bf n},{\bf n}'+{\bf e}_j^{}}^{}
+ D\,\delta_{{\bf n},{\bf n}'-{\bf e}_j^{}}^{}\right),
\eeq
where the coefficients $A - D$ have a (local) dependence on $\sigma_{x,\tau}^{}$ which is inconsequential for the
present purposes and depends of the specifics of the Hubbard-Stratonovich transformation. The boundary conditions are 
periodic in the spatial directions and anti-periodic in the time direction.

We have assumed that our system is spin-symmetric, {\it i.e.} $M \equiv M_\uparrow^{} = M_\downarrow^{}$ such that the fermion sign problem is absent,
provided that the interaction is attractive~($U < 0$). In Eq.~(\ref{HSZ}) $S_g^{}[\sigma]$ is a local action, the form of which 
depends on the specific choice of Hubbard-Stratonovich transformation. We have also assumed that a Suzuki-Trotter 
discretization~\cite{Suzuki} of the imaginary time dimension into $N_\tau^{}$ slices is performed, although it should be noted
that continuous time formulations exist~\cite{Rombouts}, and are yet to be studied in combination with the HMC algorithm.
It should also be noted that the decomposition of Eq.~(\ref{Bblocks}) can in principle be improved by using a higher order form in $\tau$.
For the sake of simplicity, we shall focus on the above formulation for the remainder of this work.


\section{Hardware \label{hardware}}

Our GPU machine is equipped with a 3.0~GHz AMD Phenom II X4 940 quad-core CPU, along with four NVIDIA Tesla C1060 GPU cards. 
However, in order to benchmark the performance of our CUDA~(GPU) implementation, we have used a 2.66~GHz Intel Xeon
E5430 CPU for the GPU/CPU comparison. In practice, we have found that the Intel Xeon CPU outperforms the AMD Phenom II CPU
by a factor of $\sim 2$. In order to avoid issues with the OpenMP parallelization of the CPU implementation, the CPU timing runs
have been performed using a single thread only.
\begin{figure}[b]
\epsfig{file=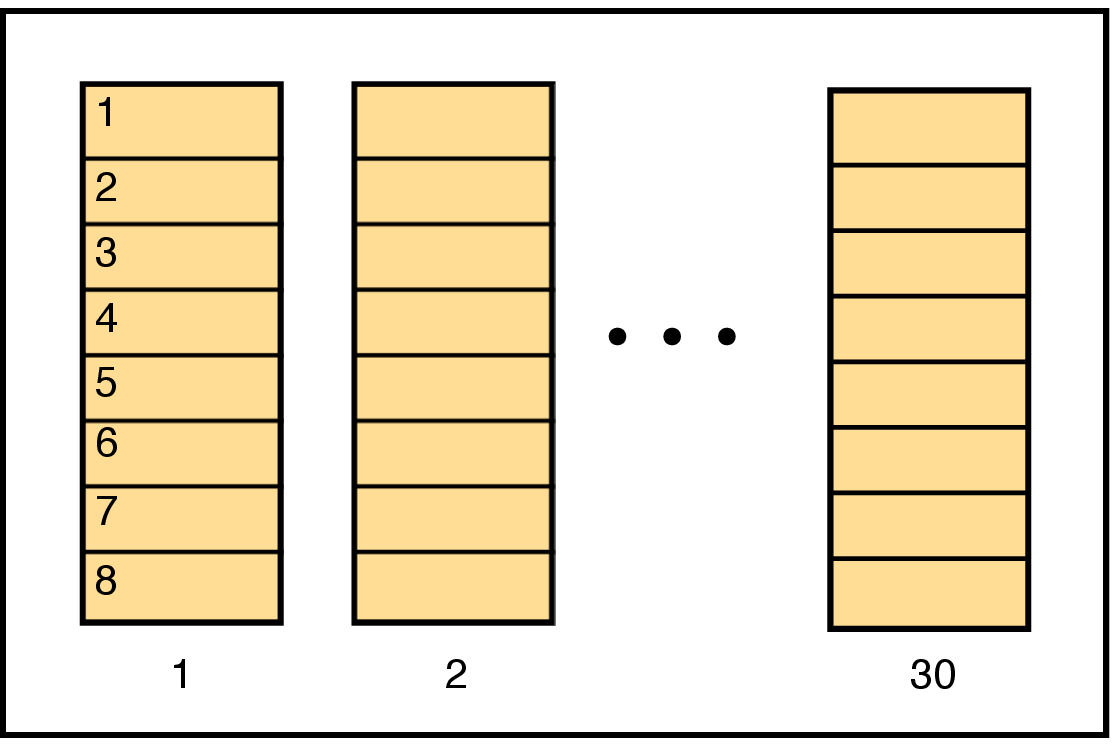, width=.36\columnwidth} 
\epsfig{file=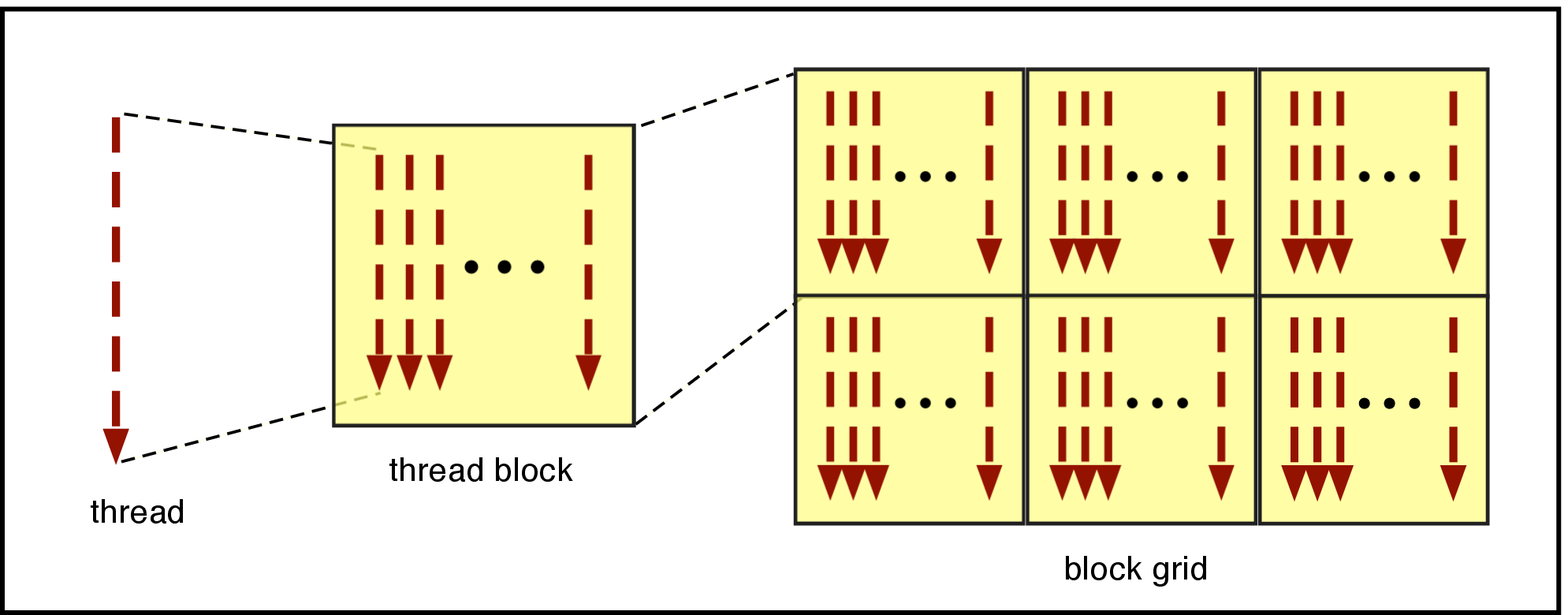, width=.62\columnwidth} 
\caption{Left panel: The Tesla C1060 GPU contains 30 multiprocessors, each equipped with 8 scalar processors, for a total of
240 processing units.
Right panel: Threads are organized in blocks that run the same instructions simultaneously but on different data. 
Blocks are organized in a grid, and are in general not executed simultaneously.}
\label{Fig:multiproc}
\end{figure}

Each C1060 GPU card consists of 30 multiprocessors, with 8 scalar 
processors each (Fig.~\ref{Fig:multiproc}, left panel). Code is executed on these processors by means of subroutines,
commonly referred to as CUDA kernels. A scalar processor executes a given kernel in parallel on a group 
of threads referred to as a thread block. Each thread in the block executes the same instructions in the kernel on different data 
(Fig.~\ref{Fig:multiproc}, right panel). Different blocks can execute different kernels on the same scalar processor, 
with up to 128 threads per scalar processor.
Each multiprocessor has 16 kB of shared memory (shared between its scalar processors), 
and approximately $2^{11}_{}$ 4-byte registers, private to each scalar processor. Memory bottlenecks aside, 
achieving maximum performance (running 1024 threads executing the same CUDA kernel simultaneously) 
requires each thread to use no more than 16 registers for single precision 
(8 for double precision)~\cite{CUDAsheet}.  Additionally, for the threads to execute in parallel, the
execution path has to be non-divergent, such that the use of conditional statements is prohibited. In practice, this restricts the
complexity of the kernels intended for execution on the GPU.


\section{Algorithm and implementation \label{algorithm}}

\subsection{Hybrid Monte Carlo}

The Hybrid Monte Carlo~(HMC) algorithm~\cite{Kennedy} has proven to be one of the most efficient approaches to 
calculations of the thermodynamic properties of many-fermion systems, and it currently forms a fundamental part of 
state-of-the-art Lattice QCD calculations. In one of its most popular forms, known as the $\varphi$-algorithm~\cite{Gottlieb}, 
HMC combines the stochastic evaluation of the fermion determinant with the molecular dynamics~(MD) evolution of the auxiliary 
Hubbard-Stratonovich field $\sigma$ (or the gauge field $A_\mu^{}$ in the case of Lattice QCD). To this end, one introduces the
pseudofermion representation 
\beq
(\det M[\sigma])^2 \:=\: \det M^T_{}\!M[\sigma] = \int \mathcal D \phi^\dagger \mathcal D \phi \; \exp(-S_{p}^{}[\sigma]),
\eeq
of the fermion determinant, such that the pseudofermion action is
\beq
S_{p}^{}[\sigma] \equiv \sum_{{\bf n},\tau} \phi^\dagger Q^{-1}_{}[\sigma] \phi, 
\eeq
with $Q \equiv M^T_{}\!M$. This approach enables global updates of the auxiliary field $\sigma$, and can be made free of systematic errors associated
with the MD integration step by means of a Metropolis accept/reject step at the end of each MD trajectory. HMC thus offers vastly improved scaling 
with system size in comparison with conventional determinantal Monte Carlo~(DMC) algorithms~\cite{DMC}, especially when an iterative solver is used
in conjunction with the MD evolution.

The fundamental component of the HMC algorithm is the calculation of the force term in the MD evolution through solution of the linear 
problem $Qx = b$. This entails the repeated inversion of the sparse, symmetric and positive-definite matrix $Q$ with a constant 
structure of non-zero elements, which change over the course of the MD evolution. This problem is frequently
quite ill-conditioned, espacially at low temperatures and in the vicinity of quantum critical points. 
One of the most robust methods to invert symmetric positive-definite matrices is the conjugate gradient~(CG) algorithm 
(see {\it e.g.} Ref.~\cite{CG}), which has become a standard workhorse in HMC simulations as it offers superior scaling over direct
methods based on LU-decomposition. When CG is used, the most time-consuming part of the HMC algorithm consists of 
repeated matrix-vector operations involving the fermion matrix $Q$. As the CG algorithm is very sensitive to the condition number of $Q$,
the computational cost of the HMC algorithm may frequently become prohibitive, unless a way is found to efficiently precondition the problem.

\subsection{Polynomial Preconditioning}

The sensitivity of the HMC algorithm to the condition number of $Q$ becomes particularly problematic, for instance, in non-relativistic 
systems at low temperatures, an example of which is the Hubbard model considered here. The same issue arises for relativistic systems 
at small fermion mass (such as Lattice Monte Carlo simulations of QCD or the low-energy theory of graphene), where it is referred to as 
the ``Berlin Wall'' problem. The sensitivity to the condition number leads to two closely related problems that always appear in HMC 
simulations to some extent. Firstly, as the fermion matrix becomes ill-conditioned, the magnitude of the fermion force in the MD evolution grows.
In order to maintain a high acceptance rate, the MD equations of motion then have to be integrated with a progressively smaller stepsize, with
a concomitant increase in computation time. In practice, the HMC algorithm typically breaks down rapidly below some characteristic temperature. 
A number of solutions exist that go a long way towards mitigating this problem, culminating in the Polynomial HMC algorithm~\cite{FrezzottiJansen}
which combines polynomial filtering and multistep integration with the standard HMC approach. While improving the acceptance rate of HMC is an active
field of study, this issue is not essential to our present objectives and will thus not be considered further.

Secondly, the CG inversion of an ill-conditioned matrix requires a large number of iterations which, apart from introducing excessive computational
requirements, may accumulate significant round-off error and thus fail to converge altogether. Mitigating this problem, and keeping the number of CG
iterations within reasonable bounds through efficient preconditioning techniques, is central to our study. A preconditioning technique which has
proven to be very effective is based on approximating $Q^{-1}_{}$ by means of a Chebyshev polynomial in $Q$, given by the product representation
\beq
P(Q) = c_{2n}^{}\prod_{k=1}^{2n} \left(Q - z_{2n,k}^{}\right),
\label{Cheb}
\eeq
where 
\beq
c^{-1}_{2n} = \frac{1+\epsilon}{2} \prod_{k=1}^{2n}\left(\frac{1+\epsilon}{2}-z_{2n,k}^{}\right),
\eeq
with the roots
\beq
z_{2n,k}^{} = \frac{1 + \epsilon}{2}\left[1 - \cos\left(\frac{2\pi k}{2n + 1}\right) \right] -
i \sqrt{\epsilon}\,\sin\left(\frac{2 \pi k}{2n + 1}\right),
\eeq
where we assume $Q$ to be normalized such that the eigenvalues are in the interval~$[0,1]$. The Chebyshev polynomials provide
an optimal approximation in the eigenvalue interval~$(\epsilon,1]$, in the sense that the relative error 
\beq
|R_{2n,\epsilon}^{}| \leq 
2\left(\frac{1-\sqrt{\epsilon}}{1+\sqrt{\epsilon}}\right)^{2n+1}_{},
\eeq
where
\beq
R_{2n,\epsilon}^{}(z) = z\left[ P_{2n,\epsilon}^{}(z) - \frac{1}{z} \right],
\eeq
is exponentially suppressed with the degree $n$ of the polynomial. An appealing feature of the Chebyshev preconditioner is
that the efficiency and computational effort can be tuned by varying the degree $n$ of the polynomial. However, care should be taken
as the product representation of Eq.~(\ref{Cheb}) is susceptible to round-off error for large $n$. This problem can be largely eliminated
by optimizing the ordering of the roots $z_{2n,k}^{}$~\cite{ChebOrdering}, without resorting to a recursive Clenshaw approach.

As both the CG inversion itself and the preconditioning step entails repeated application of the matrix $Q$, the performance of
Chebyshev preconditioned HMC calculations depend crucially on the efficiency of the matrix-vector~(MV) operation 
\beq
Y^{}_i = \sum_{j=1}^{N_x^3 \times N_\tau^{}} Q^{}_{ij} X^{}_j, 
\eeq
where $j$ is a collective index for a site in the spacetime lattice $(x,\tau)$. We thus proceed to study how this operation can be
efficiently implemented on GPU architectures, focusing on the finite-temperature Hubbard model 
in one and three spatial dimensions.

\subsection{Implementation}

The most time-consuming operation in the HMC algorithm is the application of $Q \equiv M^T_{}\!M$ to a given vector, which is typically
performed by successive application of $M^T_{}$ and $M$. The primary challenge consists of fully utilizing the scalar processors of 
the GPU (also referred to as CUDA cores), with careful consideration of the hardware limitations. The objective is to maximize the
number of simultaneously running threads, {\it i.e.} the occupancy of the GPU (expressed as a percentage of the maximum number of concurrent
threads). Because the available shared memory is limited, achieving $100\%$ occupancy is challenging for most problems. However, with 
careful memory management the occupancy can be optimized, thus yielding significant performance gains.

\begin{figure}[b] 
\centering
\epsfig{file=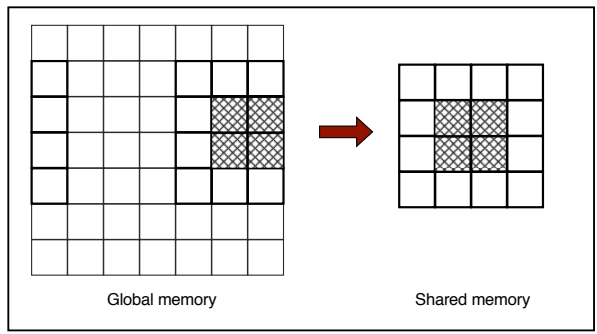, width=.47\columnwidth} 
\caption{Two-dimensional representation of the thread-blocking procedure. Doubly hatched sites are assigned to a specific block of threads.
The doubly hatched sites depend on the singly hatched ones, which are also included in shared memory. The inclusion of this ``boundary layer"
minimizes the interaction with the (slow) global memory of the GPU, as well as the overhead associated with the implementation of boundary
conditions at the lattice edges. Notice that the corners of the boundary layer are included for simplicity.}
\label{Fig:memory}
\end{figure}

The structure of Eq.~(\ref{dDM}) indicates that any given element of the resultant vector $Y$ depends only on a limited number 
of neighboring elements in $X$ and $\sigma$. This suggests that a good strategy is to assign each lattice site to one thread, such
that each thread $j$ computes the element $Y_j^{}$. Upon completion of this operation, $M^T$ is applied to 
$MX$ in a similar fashion. However, this ``naive" approach was found to yield only a very modest gain over the CPU,
in most cases not exceeding a factor of $\sim 3$. This disappointing result is due to memory bandwidth limitations 
and the intricacies of implementing periodic boundary conditions. As the element $Y_i^{}$ depends on itself and on all of its neighbors 
in the $d+1$ dimensional lattice, there are up to $2d+2$ (for a first order approximation to the Laplacian) reads to global memory for 
each lattice site ($2d$~spatial neighbors, 1 temporal neighbor, and the site itself).  Since global memory is rather slow, each read 
consumes a considerable amount of time.  

In order to overcome the memory access problems, we have implemented the ``thread-blocking" procedure illustrated in Fig.~\ref{Fig:memory}, 
whereby spatial cubes of lattice sites are saved to the shared memory of the GPU, after which each cube is assigned to a block of 
threads for computation. This reduces the number of global memory reads by a factor of~6. Additionally, each block was assigned a 
``boundary layer" according to the specified boundary conditions, meaning that checks to determine whether the lattice edge is crossed
are done once and for all, instead of separately for each lattice site. When combined, these strategies provide a speedup factor of 
$\sim 20$ or more over the naive approach. The main drawback is that the limited amount of shared memory reduces the maximum 
occupancy to $\sim 50\%$ for $d=3$. For $d=1$, the number of neighboring elements is much smaller, and thus none of the above 
issues represent a problem.

Another significant bottleneck arises due to the limited number of registers available for use by the threads in each scalar processor.
With the naive implementation, for $d=3$, the compiler allocates 64 registers for each thread, which yields a maximum occupancy of
$\sim 25\%$. By optimizing the storage of quantities in the constant memory of the GPU and carefully assigning each register, only 32 are 
needed in the end, which returns the maximum occupancy to $\sim 50\%$. Again, this problem is not encountered for $d=1$.


\section{Results and Conclusions \label{results}}

\begin{figure}[t]
\centering 
\epsfig{file=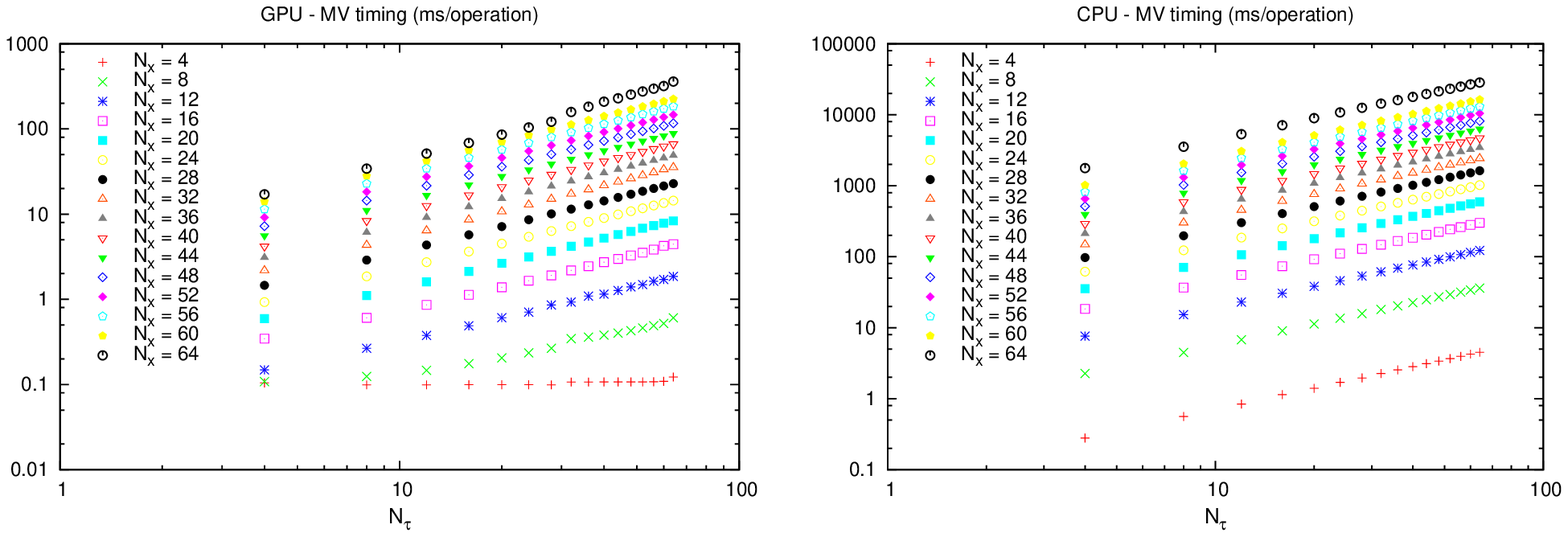, width=\columnwidth} 
\epsfig{file=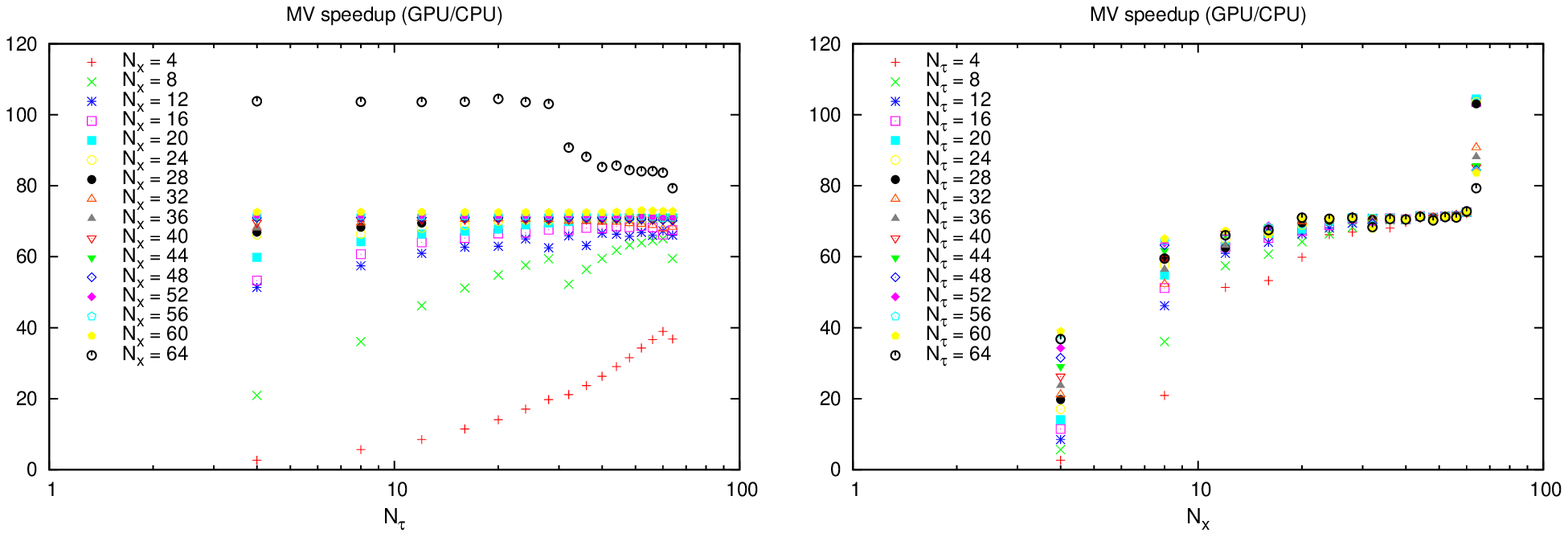, width=\columnwidth} 
\caption{Upper panels: timing of the matrix-vector~(MV) operation (in ms) for the Hubbard model in $3+1$~dimensions, as a 
function of the spatial and temporal lattice sizes~$N_x^{}$ and~$N_\tau^{}$. Upper left panel: results for the NVIDIA Tesla C1060~GPU. 
Upper right panel: results for the Intel Xeon~E5430~CPU (using~1 thread). 
Lower panels: speedup factor of the matrix-vector~(MV) operation for the Hubbard model in $3+1$~dimensions
on the NVIDIA Tesla~C1060~GPU. Lower left panel: speedup as a function of spatial lattice size~$N_x^{}$. Lower right panel: speedup as a 
function of temporal lattice size~$N_\tau^{}$. It should be noted that $\sim 94$~\% of the computational effort of the
preconditioned Conjugate Gradient~(PCG) algorithm is due to MV operations.
\label{Fig:GPU_timings}}
\end{figure}

The performance of our GPU implementation of the Hubbard model for $d = 3$ is characterized in 
Fig.~\ref{Fig:GPU_timings}, along with a determination of the speedup factor of the GPU versus CPU. 
Clearly, the GPU provides a considerable performance gain over the CPU, except at the smallest
spatial lattice sizes $N_x^{}$. At present, our Chebyshev preconditioned HMC implementation on CPU systems is limited to 
$N_x^{} = 8 - 16$, depending on temperature. Assuming the favorable (approximate) $V^{5/4}_{}$ scaling of the HMC algorithm,
the observed speedup factors of $\sim 60$ should allow for simulations to be performed at $N_x^{} > 20$ even
at low temperatures. This would be especially true for NVIDIA's new ``Fermi" architecture, where the efficiency of double precision
operations is enhanced by a factor of $\sim 5$ over the GPU architecture used in this work.

In order to obtain a realistic projected efficiency for a full HMC implementation on the GPU, we have also constructed a
GPU implementation of the full preconditioned CG algorithm. We observe that the MV operations account for $\sim 94\%$ of the total
workload, a figure which should be representative of the HMC algorithm as a whole. We conclude that Lattice Monte Carlo
simulations of the Hubbard and related models using modern GPU architectures appear to be a promising and worthwhile pursuit.


\section*{Acknowledgments}
We acknowledge support under U.S. DOE Grants No.~DE-FG02-00ER41132 and DE-AC02-05CH11231, 
UNEDF SciDAC Collaboration Grant No.~DE-FC02-07ER41457 and NSF Grant No.~PHY--0653312. 
This study was supported in part 
by the Academy of Finland through its Centers of Excellence Program (2006 - 2011), the Vilho, Yrj\"o, and 
Kalle V\"ais\"al\"a Foundation of the Finnish Academy of Science and Letters, and the Waldemar von Frenckell 
and Magnus Ehrnrooth Foundations of the Finnish Society of Sciences and Letters. Part of this work was performed 
using an allocation of computing time from the Ohio Supercomputer Center.
We thank R.~J.~Furnstahl, R.~Scalettar and N.~Trivedi for useful discussions 
and encouragement. 


\end{document}